\documentclass[aps,prb,superscriptaddress,twocolumn]{revtex4}
\usepackage{amsmath}
\usepackage{amsfonts}
\usepackage{graphicx}
\usepackage{color}
\usepackage{dsfont}

\begin{document}

\newcommand{\beq}{\begin{equation}}
\newcommand{\eeq}{\end{equation}}
\newcommand{\beqa}{\begin{eqnarray}}
\newcommand{\eeqa}{\end{eqnarray}}
\newcommand{\ben}{\begin{enumerate}}
\newcommand{\een}{\end{enumerate}}
\newcommand{\hs}{\hspace{1.5mm}}
\newcommand{\vs}{\vspace{0.5cm}}
\newcommand{\note}[1]{{\color{red} \bf [#1]}}
\newcommand{\ket}[1]{|#1 \rangle}
\newcommand{\bra}[1]{\langle #1|}

\title{Many-Body Effects in Topological Kondo Insulators }

\author{Jason Iaconis}
\affiliation{Department of Physics, University of California, Santa Barbara, CA, 93106-9530}
\author{Leon Balents}
\affiliation{Kavli Institute for Theoretical Physics, University of California, Santa Barbara, CA 93106-4030, U.S.A.
}

\date{\today}

\begin{abstract}
We study the effect of interactions on the properties of a model 2D topological Kondo insulator phase.
  Loosely motivated by recent proposals where graphene is hybridized with impurity bands from heavy adatoms with partially filled d-shells,  we introduce a model Hamiltonian which we believe captures the essential physics of the different competing phases. We show that there are generically three possible phases with different combinations of Kondo screening and magnetic order. Perhaps the most dramatic example of many-body physics in symmetry protected topological phases is the existence of the exotic edge states.  We demonstrate that our mean field model contains a region with a time-reversal invariant bulk phase but where TR symmetry is spontaneously broken at the edge. Such a phase would not be possible in a non-interacting model. We also comment on the stability of this phase beyond mean field theory.  \end{abstract}

\maketitle

\section{Introduction}
\label{sec:introduction}

The discovery of topological insulators in ``inverted'' semiconductors with strong spin orbit coupling led to an explosion of interest in topology in condensed matter systems.  The combination of topology and strong correlations of electrons is a subject of major current activity, and systems in which both are at play suspected to host a variety of unique phenomena.\cite{doi:10.1146/annurev-conmatphys-020911-125138} ``Heavy fermion'' and related materials with heavy lanthanide elements are natural places to seek such phenomena, as the electrons on these atoms experience strong spin-orbit coupling (SOC) -- which is a driving force for non-trivial topology in many systems -- and are strongly correlated.  Theory recently suggested a concrete role for topology in these systems in the form of Topological Kondo Insulators (TKIs) \cite{PhysRevB.85.045130,PhysRevLett.104.106408}.  A TKI is a Kondo lattice system involving rare earth ions which at low energies hybridize with lighter conduction electrons forming a topological insulator.  SmB$_6$ is strongly suspected to be such a TKI \cite{PhysRevB.88.180405, PhysRevX.3.011011,cite-key} .

While SmB$_6$ and indeed any Kondo lattice system is indisputably a strongly correlated electronic system, the {\em low energy} description of a TKI in terms of effective bands seems indistinguishable from that of an uncorrelated topological insulator.  This is disappointing in view of the hope for new phenomena in correlated strong SOC systems.  The aim of this article is to consider the possibility of competing states, and seek out new effects arising from electronic correlations in a system which may host a TKI.  We do this in the context of a model motivated by the recent proposal of realizing a TKI in graphene doped with heavy adatoms.  In this model, described in detail in Sec.~\ref{sec:model}, we obtain a zero temperature phase diagram which embeds the topological insulator physics into the classic Doniach phase diagram \cite{Doniach1977231} for Kondo lattice systems, with its magnetically ordered and Kondo screened phases.
The magnetic ordering quantum phase transition is discussed in this context.  We also consider the characteristic behavior of the boundary of the TKI, and show that this model is prone to {\em surface magnetic ordering}, even within the TKI state.  Such spontaneous and intrinsic  time-reversal breaking at the surface of a TKI could be the desired hallmark of correlations in the TKI state.

\subsection{The Model}
\label{sec:model}

We study a tight-binding model of graphene, with a localized $d$-orbital electron site at the center of each face on the honeycomb lattice.  Such a model was studied extensively in [\onlinecite{PhysRevLett.109.266801,PhysRevX.1.021001,PhysRevB.82.161414,PhysRevLett.108.056802}], via a combination of first-principle calculations on a tight binding model and density functional calculations.  There it was shown that the strong onsite spin-orbit term for the localized $d$-electrons, when hybridized with the conduction electrons, $c$, conspire to create a band insulator with a nontrivial topology. This topological phase on graphene is reminiscent of the original proposal of topological order in graphene by Kane and Mele in 2005 \cite{PhysRevLett.95.226801, PhysRevLett.95.146802}. Our goal is to study explicitly the effect of interactions on such a model.

In [\onlinecite{PhysRevLett.109.266801}], the DFT calculations show that the most important angular momentum states of the $d$ electron are those with $L^z$ quantum number $m = \pm 1$.  These arise from the $d_{xz}$ and $d_{yz}$ adatom states, and so we restrict our model to states with these angular momenta.  We include a spin-orbit coupling term for the $d$-electrons, but not for the conduction electrons where the small magnetic moments are expected to lead to a negligibly small amount of spin-orbit coupling.  The chemical potential of both the $c$ and $d$ electrons is set so that there are two $c$ and two $d$ electrons per unit cell.  This is a necessary condition if the hybridization is to lead to a band insulator.

The complication comes when we include an interaction term between the local $d$ moments and the nearby conduction electrons.  In order to write down this interaction, need to know the spin state formed by the composite two-body state sitting at each $d$-site. We assume that within the $m=\pm 1$ subspace, a Hund's rule type coupling makes it energetically unfavorable for both $d$-electrons to have the same angular momentum quantum number.  We therefore ignore the possibility of the $d$-electrons forming spin-2 states, $\ket{+\uparrow, + \downarrow}$ and $\ket{-\uparrow,-\downarrow}$, when writing the form of the interaction.   The remaining states are
\beqa
\vec{S}_{\text{tot}}=0 &:&  \ket{+\uparrow,-\downarrow} - \ket{+\downarrow,-\uparrow}  \nonumber\\
\vec{S}_{\text{tot}}=1 &:& \ket{+\uparrow, -\uparrow} \nonumber\\
&&  \ket{+\uparrow,-\downarrow}+ \ket{+\downarrow,-\uparrow} \nonumber\\
&& \ket{+\downarrow,-\downarrow} \nonumber.
\eeqa
Within this subspace the Kondo interaction is a spin-spin interaction which occurs between the $\vec{S}_{\text{tot}}=1$ states, while the S.O. coupling term favors the  $\ket{+\downarrow,-\uparrow}$ state, which is a linear combination of the singlet and triplet $S^z=0$ states.

We define the operator $C_{i,p,\sigma}$, which is a linear combination of $c_{i\sigma}$ operators which carry angular momentum $p= \pm 1$.
\beqa
C_{R,p,\sigma} &=& \frac{1}{\sqrt{6}} \sum_{j=1}^6 e^{-i(\pi/3) p(j-1)} c_{R + e_j,\sigma}  \nonumber \\
&=& \sum_k \left (V_p(k) c_{A,k,\sigma} + V^*_{-p}(k) c_{B,k,\sigma} \right )
\eeqa
where $A$ and $B$ denote the two sublattices of the honeycomb lattice, and $e_j$ are the nearest-neighbor lattice vectors connecting the $c$ and $d$ sites.

We can now construct a spin-1 operator for both the $c$ and $d$ electrons from two spin-$\frac{1}{2}$ operators as follows,
\beqa
\vec{S}_1 = \frac{1}{2}\sum_{m=\pm 1 }  d^\dagger_{m,\alpha} \vec{\sigma}_{\alpha \beta} d_{m,\beta} \\
s_1 = \frac{1}{2} \sum_{m=\pm 1 }  C^\dagger_{m,\alpha} \vec{\sigma}_{\alpha \beta} C_{m,\beta},
\eeqa 
where $\vec{\sigma}$ is the usual spin-1/2 Pauli vector.  The Kondo interaction is then an antiferromagnetic spin-spin interaction between the spin-1 particle on the d-sites and the effective spin-1 formed by a linear combination of the conduction electrons near this d-site. The 4 fermion Kondo interaction is therefore
\beqa
H_{K} &=& J \sum_i \vec{S}_{1,i} \cdot \vec{s}_{1,i} .\label{KondoInt}
\eeqa

We also include an RKKY spin-spin interaction in our model.  This is an interaction term between the $d$-electron states on different sites which is mediated through conduction electrons. Performing second order perturbation theory in the Kondo interaction, the $d$-spin on one site interacts antiferromagetically with the conduction electrons near that site which in turn act antiferromagetically with a neighboring $d$-site.  We assume this creates an effective ferromagnetic interaction between the two $d$-sites and parameterize this term with the variable $J_m$.   Although in principle the RKKY interaction is included in Eq.~(\ref{KondoInt}), it is difficult to derive both the Kondo and RKKY effects together\cite{PhysRevB.67.064417, PhysRevB.56.11820}. Therefore,  in our model we include as a separate parameter the RKKY interaction
\beq
H_{RKKY} = J_m \sum_{\langle i j \rangle} \vec{S}_{1,i} \cdot \vec{S}_{1,j}.
\eeq

Therefore, the simplest interacting model that we believe captures the essential many-body physics of the TKI system is\beqa \label{Ham}
H &=& \sum_{k,\sigma} \epsilon_k c_k^\dagger c_k + \lambda \sum_{i,p,\sigma} (d_{ip\sigma}^\dagger d_{ip\sigma} - 2) + \mu \sum_{i,\sigma} (c^\dagger_{i\sigma} c_{i\sigma} - 2)  \nonumber\\
 &&+ \Lambda \sum_i (d_{+\alpha}^\dagger \sigma^z_{\alpha \beta} d_{+\beta} - d_{-\alpha}^\dagger \sigma^z_{\alpha \beta} d_{-\beta}) \nonumber \\
&& + J \sum_i \vec{S}_{1,i} \cdot \vec{s}_{1,i} +  J_m \sum_{\langle i j \rangle} \vec{S}_{1,i} \cdot \vec{S}_{1,j}
\eeqa
The goal of this paper is to examine at the mean field level the effects of the two interaction terms.

\section{MFT and the Bulk Phase Diagram}
\label{sec:model-1}

\subsection{Outline}
\label{sec:outline}

In this section we will describe the general phases which exist in our model.  This allows us to embed the TKI phase within the standard phase diagram for the Kondo lattice \cite{PhysRevB.20.1969}.  We show that there is in general a first order phase transition from a phase with no Kondo order (where the d-spins order magnetically)
 to a TKI phase at some nonzero value of the Kondo exchange $J$. Furthermore, this phase can be destroyed by RKKY type spin-spin interactions
 which are present in the fully interacting theory and so we include as a separate term in our mean-field Hamiltonian. 
 If the system orders magnetically, time-reversal symmetry is broken and any distinction between the topological phase and a trivial insulator loses meaning.  A similar MF phase diagram was calculated in Ref.'s~[\onlinecite{PhysRevB.85.125128},\onlinecite{PhysRevB.89.085110}] for 3D TKIs, which did not consider the role of RKKY interactions.

\subsection{Methods}
\label{sec:methods}

We would like the study our fully interacting model at the mean field level.  This means we should decouple our Kondo and RKKY interaction terms into the appropriate channels.  This amounts to an assumption, which must be checked self-consistently, about the type of order in the interacting groundstate.  To this end, we assume that the Kondo interaction, 
\beq
\vec{S}_i \cdot \vec{s}_i= \sum_{m,m^\prime} \vec{\sigma}_{\alpha\beta} \cdot \vec{\sigma}_{\gamma \delta} C^\dagger_{m,\alpha} C_{m,\beta}d^\dagger_{m^\prime,\gamma} d_{m^\prime,\delta}
\eeq
induces a nonzero expectation value for the operator $\langle C^\dagger_{i,p,\sigma} d_{i,p,\sigma}\rangle$, and the RKKY interaction causes a nonzero expectation value for the magnetic order parameters $\langle \vec{S}_i \rangle$.

We can now perform the standard mean-field analysis by decoupling the interactions into the channels
\beqa
&&J \, C^\dagger_\alpha d_\alpha d^\dagger_\beta C_\beta \hs  \rightarrow \hs  J ( \phi_{\alpha} d^\dagger_\beta C_\beta + \phi_\beta^* C^\dagger_\alpha d_\alpha) - J\phi_\alpha \phi^*_\beta. \nonumber\\
&& J_m(S^xS^x + S^yS^y) \hs \rightarrow \hs 2J_m \langle S^x \rangle S^x - J_m\langle S^x \rangle^2 
\eeqa
Where $\alpha$ and $\beta$ denote the two types spin-orbit coupled single particle states.  

Notice that the interaction in Eq.~\eqref{KondoInt} contains terms like $C^\dagger_+ d_- C_+ d^\dagger_-$, but these terms vanish after the above substitution since the resulting $c$-$d$ hopping process does not conserve angular momentum.  Also, we assumed that the spin-spin interaction favors magnetic order in the XY plane. We will justify this assumption in the next section.

We are therefore left with a non-interacting Hamiltonian 
for which the parameters $\phi_+$, $\phi_-$ and $\langle S^x \rangle$ must be determined self-consistently. We use the parameters
 $\mu$ and $\lambda$ in Eq.\eqref{Ham} to enforce the conditions that there are 2 electrons per d-site and one electron per graphene site, 
ensuring that the system is a band insulator when there is a full gap at the Fermi energy.

Performing these substitutions, the MF Hamiltonian becomes

\beqa \label{MFHam}
H &=& \sum_{k,\sigma} \epsilon_k c_k^\dagger c_k + \lambda \sum_{i,p,\sigma} (d_{ip\sigma}^\dagger d_{ip\sigma} - 2) + \mu \sum_{i,\sigma} (c^\dagger_{i\sigma} c_{i\sigma} - 1)  \nonumber\\
 &&+ \Lambda \sum_i (d_{+\alpha}^\dagger \sigma^z_{\alpha \beta} d_{+\beta} - d_{-\alpha}^\dagger \sigma^z_{\alpha \beta} d_{-\beta}) \nonumber \\ \nonumber\\
&&-  J \sum_i\sum_{p,\sigma}(\phi_+ + \phi_-)( C^\dagger_{i p\sigma} d_{ip\sigma}+  \text{h.c.})   \nonumber\\
 &&- 2J_m \sum_{i} \langle S^x \rangle (d^\dagger_{+\uparrow} d_{+\downarrow} + d^\dagger_{-\uparrow}d_{-\downarrow} + \text{h.c.}) \nonumber \\
&& + J ( \phi_+  + \phi_-)^2 + J_m \langle S^x\rangle^2
\eeqa

We then diagonalize the mean-field Hamiltonian and write the bare electron operators in terms of the free excitations of the system 
\beqa
d_{k,\alpha \sigma}^\dagger = \sum_n \alpha^{(\alpha \sigma)}_{k,n} a_{k,n}^\dagger \nonumber\\
c_{k,\alpha \sigma}^\dagger = \sum_n \beta^{(\alpha \sigma)}_{k,n} a_{k,n}^\dagger,
\eeqa  
which are linear combinations of $c$ and $d$ electrons.  Since $H$ is diagonal in the $a_n^
\dagger$ operators, knowledge of the coefficients $\alpha_n$ and $\beta_n$ allow us to numerically determine the values of $\langle C_\pm^\dagger d_\pm \rangle$, $\langle S \rangle$, $\langle d_i^\dagger d_i \rangle$ and $\langle c_i^\dagger c_i \rangle$.  By averaging over all occupied states of the mean-field model we can then find the parameters of $H$ where the mean field constraints are satisfied.

In general, for each order parameter, $\langle \hat{X} \rangle$ there are two self-consistent solutions.  These are $\langle \hat{X} \rangle = 0$ and $\langle \hat{X}\rangle = x \neq 0$.  The solution with a nonzero order parameter breaks a symmetry of the interacting Hamiltonian and is therefore a distinct phase from the case where $\langle \hat{X} \rangle = 0$.  Furthermore, notice that the mean field Hamiltonian $H=H^{0}(J_{_X} x)$ is only a function of the product $J_{_X}$ and $x$.  Therefore, for any value of $x$ there will always be a value of $J_{_X}$ such that $\langle X \rangle = x $.  At this $J_{_X}$ there are two solutions to the mean-field conditions, $x=\langle X\rangle \neq 0$ and $x=0$.  

The mean field solution is equivalent to the lowest energy noninteracting variational state $\ket{x}$. $E_{var} = \bra{x}\mathcal{H} \ket{x}$. $E_{var}$ is a local minimum for the eigenstate $\ket{x}$ of $H^{MF}$ when the self consistency equation $x=\langle X \rangle$ is satisfied. In this case $\bra{x}\mathcal{H}\ket{x} =  \bra{x} H^{MF} \ket{x}$.  This argument allows us to compare the energies of the different self consistent solutions. Since these energy of solutions equals the variational energy, the solution with the lowest energy must be the best mean field approximation to the true groundstate.

If the order-disorder transition is continuous, then every solution with nonzero $x$ must be of lower energy than the disordered solution where $x=0$.  However, near a first order transition, there will exist solutions with small $\langle X \rangle$ that are of higher energy than the disordered state $\langle X \rangle =0$.  This gives us an easy way to distinguish between the two types of phase transitions in our calculation.

Note that model (\ref{MFHam}) does not conserve particle hole symmetry, whereas the fully interacting model (\ref{Ham}) does, so that in order to ensure there are the same number of conduction electrons we need to adjust $\mu$ so that each site on the graphene lattice is filled to $n_c = 2$.

\subsection{Phases}
\label{sec:phases}

There are three distinct phases in model $\eqref{MFHam}$, which occur when {\bf a)} $\langle C^\dagger d\rangle = 0$, {\bf b)} $\langle C^\dagger d \rangle \neq 0$ and $\langle S \rangle = 0$ or  {\bf c)} $\langle C^\dagger d \rangle \neq 0$ and $\langle S \rangle \neq 0$. 

The first case occurs when $\langle C_+^\dagger d_+ \rangle = \langle C_-^\dagger d_- \rangle = 0$ and there is no Kondo screening of the $d$-electrons.  This is the ``Fully Polarized'' or Magnetic phase. This phase may be considered uninteresting as there is no interplay between the $c$ and $d$ electrons, which know nothing about each other at the mean field level.  The conduction electrons are thus completely noninteracting and form a semimetal exactly like graphene with a pair of gapless Dirac cones in the band structure.  The $d$-electrons still contain the spin-orbit term and interact through the spin-spin interaction $J_m$.  In principle the spins could order in any number of phases depending on the exact form of the Heisenberg interaction, however the most likely result is that they order magnetically. 

Now, if the spins ordered magnetically in the $z$-direction, the S.O. term favors the state $\ket{+\downarrow,-\uparrow}$, while the $J_m$ term favors the state $\ket{+\uparrow,-\uparrow}$.  These states are completely orthogonal, therefore the spins will order in one of these two states depending on which interaction is stronger.  If $\Lambda > J_m$, the first state will be chosen and the energy per site is $E_\Lambda = -\Lambda$.  On the other hand if $J_m > \Lambda$, the spins will align in the $S^z =+1$ state and the energy per spin will be $E_{Z} = -J_m$.  On the other hand, if the spin order in the $XY$ plane, the spins order depending on the Hamiltonian
\beqa
&&H_d = \vec{v}^\dagger \left [ \begin{array}{cccc}
\Lambda &J_m S^x&0&0\\
J_m S^x&-\Lambda&0&0\\
0&0&-\Lambda&J_m S^x\\
0&0&J_m S^x&\Lambda
\end{array} \right] \vec{v}  \nonumber\\
&&\text{where   } \hs \vec{v} = (\begin{array}{cccc}d_{+\uparrow} d_{+ \downarrow} d_{- \uparrow} d_{- \downarrow} \end{array}) . \label{magHam}
\eeqa 
The energy of the groundstate with $n_d = 2$ is $E_{XY} = -\sqrt{\Lambda^2 + (J_m S^x)}^2$, where the self-consistent value of $S^x = \langle S^x \rangle$ depends on the ratio $J_m/\Lambda$.  Clearly, by the variational principle, $E_{XY}\le E_{Z}$ and $E_{XY} \le E_\Lambda$ for all $\Lambda$ and $J_m$.  So ordering in the $XY$ plane allows the spins to partially satisfy both the S.O. and {\em RKKY} terms in the Hamiltonian, so that we can safely assume that the spins order this way.  Therefore, in the fully polarized phase the $c$ and $d$ electrons behave independently, with the $d$ electrons ordering magnetically according to the Hamiltonian in (\ref{magHam})

The second phase we consider occurs when $\langle C_\pm^\dagger d_\pm \rangle \neq 0$ and $\langle S \rangle = 0$.  We will call this phase the topological Kondo insulator (TKI), or simply the Kondo phase. The nonzero value of $\phi$ causes the $c$ and $d$ electron states to hybridize.  Since $\langle S \rangle = 0$, the RKKY interaction has no effect on the mean-field groundstate. This regime can be thought of as the many-body analog of the single impurity Kondo problem.  In that problem when a single magnetic impurity is immersed in a metal, at low enough temperatures the magnetic moment is screened via the formation of a singlet pair with the nearby conduction electrons.  In our model, there is a dense lattice of magnetic moments, and every moment is completely screened by the nearby graphene conduction electrons. In ref.~[\onlinecite{PhysRevB.30.3841}], it was shown that such a phase is the stable solution to the Kondo lattice problem in the large $N$ limit, and that in this limit the phase is indeed equivalent to a conduction sea screening a dilute set of magnetic impurities.  The result of the screening at each lattice site is that the conduction electrons and local moments hybridize to form a single composite object. Therefore the Kondo phase is a free electron system where the resulting free electrons are linear combinations of the bare $c$ and $d$ electrons and are the eigenstates of $\eqref{MFHam}$ with $\langle S^x \rangle = 0$.  

The hybridization between the $c$ and $d$ electrons opens a gap at the $K$ and $K^\prime$ points in the graphene band structure (that is at the gapless nodes in the Brillouin zone). However, the form of the coefficient, $V_\pm(k)$, ensures that there is no hybridization at $\vec{k}=0$.  Therefore, when $\Lambda = 0$, the $d$-electron like band in the band structure remains gapless at $\vec{k}=0$.  However, a nonzero $\Lambda$ opens a gap at the B.Z. center and creates a full band gap.  The presence of the spin-orbit coupling also gives the band structure a nontrivial topology.  This can be seen by calculating the parity eigenvalues at the special time-reversal invariant momenta in the B.Z..  According to ref.~[\onlinecite{PhysRevB.76.045302}], a topological invariant $\nu$ can be defined by
\beq
(-1)^\nu = \prod_{j=1}^{8} P(k_j)
\eeq
whereby $\nu=0$ corresponds to the topologically trivial phase and $\nu=1$ the nontrivial phase.  The parity eigenvalues can be calculated directly by diagonalizing the mean-field Hamiltonian.  This was done in Ref.~[\onlinecite{PhysRevLett.109.266801}] and we have repeated this calculation, verifying that the TKI phase is indeed topologically nontrivial.  We have also solved the model self-consistently on a finite width strip with armchair edges, and have shown that in the Kondo phase when $\Lambda \neq 0$, at every energy within the band gap there exists a single Kramers pair of gapless edge states.  In section III, we will study the effect of the RKKY interaction on these edge states and comment on the possibility there exists a phase where time-reversal symmetry is broken on the edge while preserved in the bulk.

The final, distinct, phase that can occur is one where there is incomplete Kondo screening of the $d$ moments, so that there sis a mixture of magnetic order and Kondo order. It is important to note the previous Kondo phase is only a mean-field state and that the fully interacting many-body phase contain additional correlations. In our model we deal with these correlation by including the RKKY term in our Hamiltonian.  That is, both $\phi = \langle C^\dagger d \rangle$ and $\langle S \rangle$ are nonzero in this phase. In this case, time-reversal symmetry is broken everywhere in the bulk. In the presence of such spontaneous symmetry breaking there is no distinction between the trivial band insulator and the topological insulator phases.

\subsection{Phase Diagram and Transitions}
\label{sec:phase-diagr-trans}

\begin{figure*}[t]
a) \includegraphics[width=1.02\columnwidth]{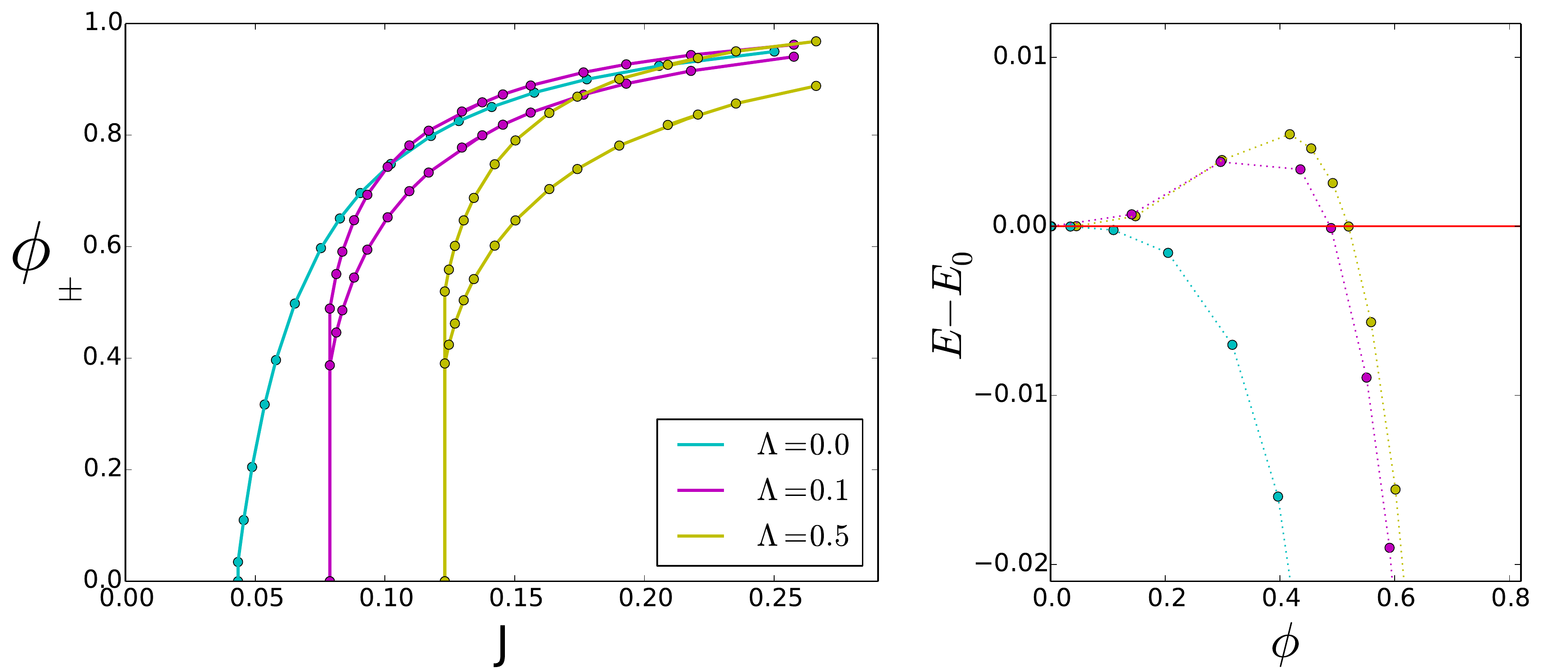}
\hspace{0.5mm} b) \includegraphics[width=0.93\columnwidth]{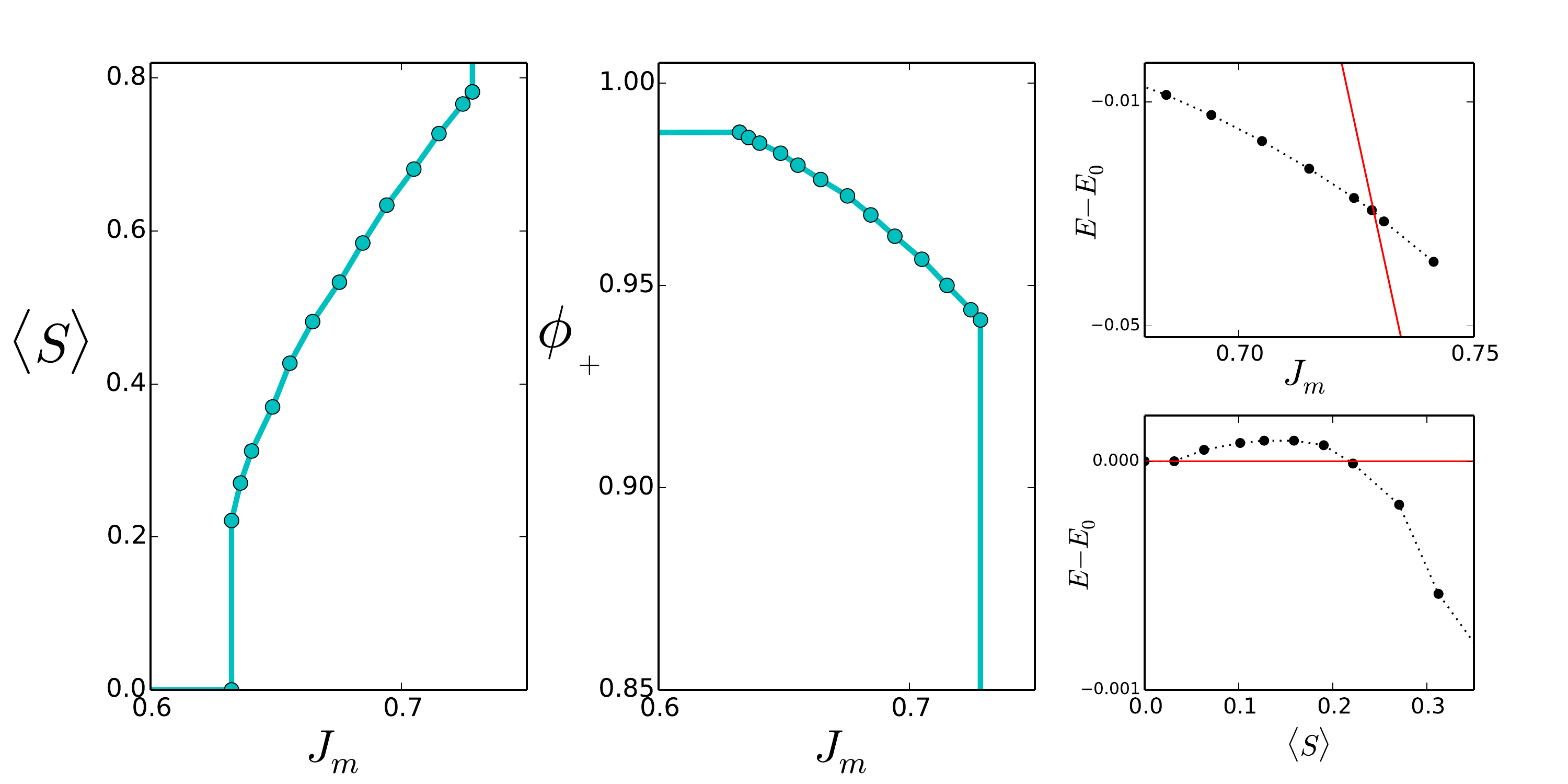}
\caption{{\bf{a)}} The Kondo order parameter for three values of $\Lambda$, ranging from $\Lambda=0$ to $\Lambda = 0.5$ {\it (left)} and the corresponding energy gain compared to the $\phi=0$ solution {\it (right)} . When $\Lambda=0$ the transition is continuous but becomes first order when $\Lambda \neq 0$.{ \bf b)} The magnetic {\it (left)} and Kondo {\it (middle)} order parameters for fixed $J=0.3$, $\Lambda=0.5$. The onset of magnetic order with increasing $J_m$ shows the transition from the Kondo phase into the mixed phase.  Further increasing $J_m$ to the point where the Kondo order parameter drops to zero this signals the transition into the fully polarized phase.  The energy crossover of this mixed phase with the fully magnetic {\it (top-right)} and TKI {\it (bottom-right)} phases show that both these transitions are first order. 
\label{order1}}
\end{figure*}

In this section we give the results of our mean-field calculation for the bulk system. 
\vspace{1pc}

\noindent \emph{Kondo Order--} 
We start with the simplest case, where we set $J_m=0$ so that spin-spin interactions do not  compete with the tendency for Kondo order.  However, even this simplest case shows interesting physics.  First studied by Withoff and Fradkin in Ref.~[\onlinecite{PhysRevLett.64.1835}], who showed that the effect of a vanishing density of states in the band structure creates a critical point $J_c$ below which the Kondo effect does not take place. Careful renormalization group calculations on the spin-$\frac{1}{2}$ pseudogap Kondo model \cite{0034-4885-76-3-032501} have since shown that, in the case of graphene, there are significant corrections to the large-N result near such a critical point and that these results depend on the presence or absence of particle-hole symmetry in the model \cite{PhysRevB.57.14254, PhysRevB.70.094502, 0034-4885-76-3-032501}.  

Our mean field calculation, however, is similar to other large-N studies of the pseudogap Kondo problem \cite{PhysRevLett.86.296}. Here the large-N mean field approach is used in order to accurately describe the physics within the Kondo phase, at the cost of not being able to accurately describe the critical point of this phase transition.  In the following we will describe the complete solution of the mean-field calculation, including all phase transitions.  We should keep in mind, however, that the quantitative details of any continuous transition are valid only as $N\rightarrow \infty$ and are not expected to hold beyond the mean field level.

When $\Lambda = 0$, all $d$-electron sites are equivalent and there is only a single order parameter $\langle C^\dagger_+ d_+\rangle = \langle C^\dagger_- d_-\rangle = \phi$.  In Ref.~[\onlinecite{PhysRevLett.64.1835}]  it was shown that the effect of a non-constant density of states leads to a critical $J_c=\frac{1}{\rho_0 D}$, which signals the onset of Kondo order in this model.

We will see how this works in the mean field calculation of our graphene model. Consider the Lagrangian form of Eq.~\eqref{MFHam},
\beq
\mathcal{L} = \int d\omega dk \left [ v^\dagger_\alpha (\mathcal{H}_{\alpha \beta}^{\text{MF}}(k) - i \omega \delta_{\alpha \beta})v_\beta  + I_\alpha v_\alpha +v^\dagger_\alpha \bar{I}_\alpha \right ].
\eeq
Integrating out the fermions, $\vec{v} = (c_A, c_B , d_+, d_-)$, produces the generating function,
\beq
\mathcal{Z} = \exp\left[\int  \frac{d\omega}{2\pi} dk \hspace{2mm} \bar{I}_\alpha \bigg(\mathcal{H}_{\alpha \beta}^{\text{MF}}(k) - i \omega \delta_{\alpha \beta}\bigg)^{-1} I_\beta \right]. \label{partitionfunction}
\eeq
The desired correlation functions can then be derived from the generating function using the expression
\beq
\langle v_{\alpha}^\dagger v_{\beta} \rangle = \int dk d \omega  \, \left (\mathcal{G}_{\alpha \beta}(k,i \omega) \right ) = \left. \frac{\delta^2 \mathcal{Z}}{\delta \bar{I}_\alpha \delta I_\beta}\right |_{I=0}
\eeq
This is analytically tractable for small values of $\phi_\pm$, where we can easily take the inverse in Eq.~(\ref{partitionfunction}) and throw out all terms of $\mathcal{O}(\phi^3)$ or greater.  This gives
\begin{widetext}
\beqa
\langle C_+^\dagger d_+ \rangle &=&\int dk d\omega \frac{J(\phi_+ + \phi_-)(V_+e^{i \theta} -V_-^*)} {(\lambda + \Lambda - i \omega)(-|f(k)| - i \omega)} \nonumber \\
\langle C_-^\dagger d_- \rangle &=& \int dk d\omega \frac{J(\phi_+ + \phi_-)(V_-e^{i \theta} -V_+^*)} {(\lambda - \Lambda - i \omega)(-|f(k)| - i \omega)} \nonumber\\
\langle d^\dagger_{\pm}d_\pm \rangle &=& \int dk \int d\omega \frac{(|f|-i \omega)(-|f|-i\omega)}{(\lambda \pm \Lambda - i \omega)(|f|-i\omega)(-|f| -i \omega) + b_1\phi^2 (-|f|-i\omega) + b_2 \phi^2 (|f|-i\omega)}
\eeqa
\end{widetext}
 \nonumber
where we have written the graphene dispersion as $f(k) = |f(k)|e^{i \theta(k)}$.  We perform the $\omega$ integral by contour integration and exchange the momentum integral for an integral over energy, $\int k dk = \int_0^D \rho_0 \epsilon d\epsilon$, where $D$ is the bandwidth and $\rho_0$ is the density of states near the Fermi energy.

The conditions $\langle C^\dagger_+ d_+ \rangle = \phi_+$, $\langle C^\dagger_- d_- \rangle = \phi_-$ and $\langle d^\dagger d \rangle = 2$ lead to the three equations
\beqa
&&(\lambda - \Lambda) \log \left[ \frac{\lambda - \Lambda}{D+\lambda-\Lambda}\right] \sim -D + \frac{1}{\rho_0 J(1 + \phi_+/\phi_-)} \nonumber\\
&&(\lambda +\Lambda) \log \left[ \frac{\lambda + \Lambda}{D+\lambda+\Lambda}\right] \sim -D + \frac{1}{\rho_0 J(1 + \phi_-/\phi_+)} \nonumber \\
&& \phi^2 \sim \lambda -\Lambda \label{analyticEqs}
\eeqa
where we ignored the constant factor from the $k$ dependence of $V_\pm$ and $\theta$, and the third relation comes from looking at the shift in the residue $i \omega$ due to small but nonzero values of $\lambda$ and $\phi$.

When $\Lambda = 0$, then $\phi_+/\phi_- = 1$. In this case there only exists a solution for $\Lambda$ when $J>\frac{1}{\rho_0 D}$.  Near $J_c = \frac{1}{\rho_0 D}$, Eq's~(\ref{analyticEqs}) give the solution $\lambda = J-J_c$.  Therefore, $\phi = \langle C^\dagger d \rangle = \sqrt{J-J_c}$, and there is a second order transition at $J = J_c$.

 When there is a nonzero spin-orbit coupling $\Lambda \neq 0$, the ratio $\phi_+/\phi_- \neq 1$, changing the solution to the first two equations in (\ref{analyticEqs}). Therefore, the value of the critical $J$ for one of these equations is increased while the other decreases. Since both equations must be satisfied the overall effect of $\Lambda$ is to push the transition back to larger values of $J_c(\Lambda)>J_c(0)$.

\begin{figure*}[t]
{\bf a)} \includegraphics[width=0.3\textwidth]{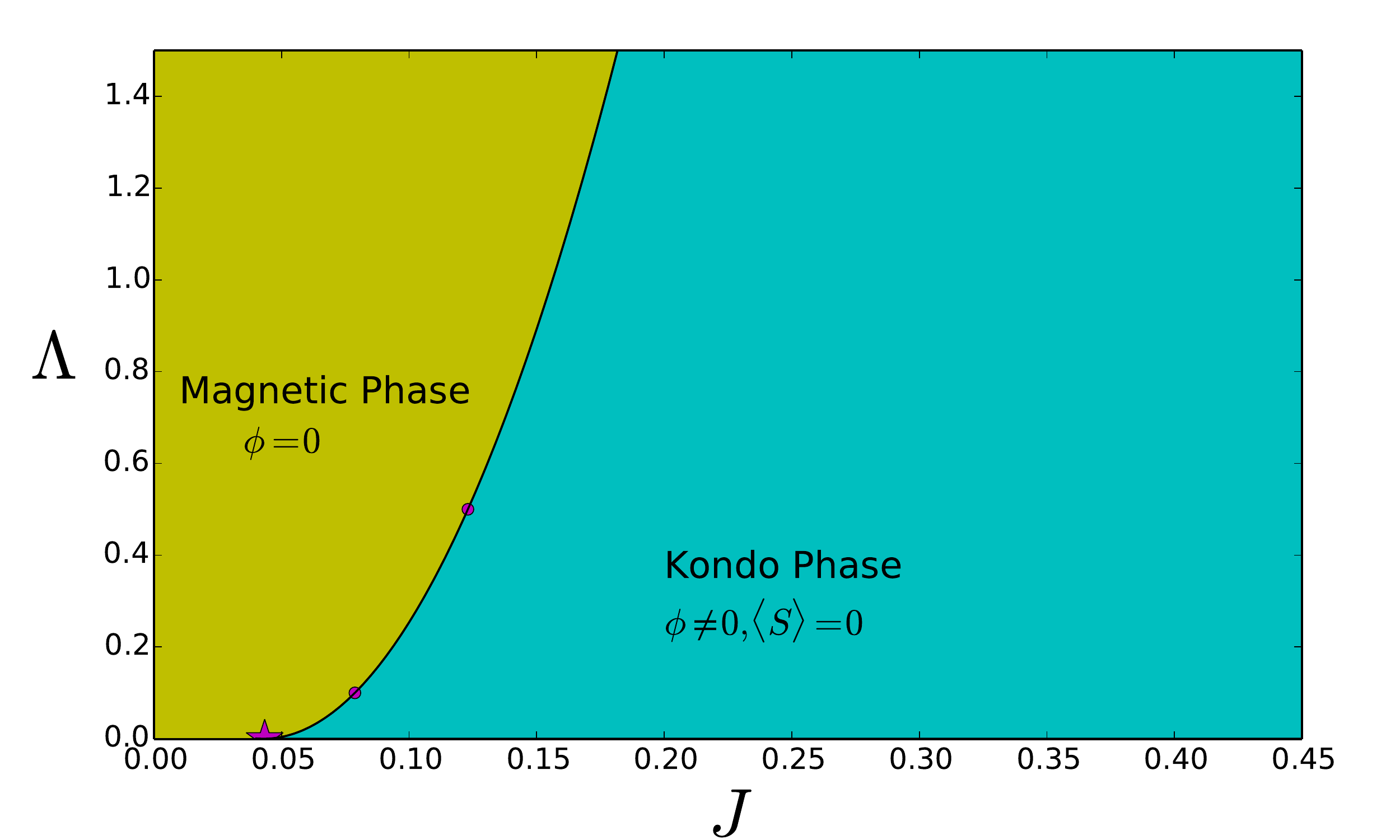}
{\bf b)}\includegraphics[width=0.30\textwidth]{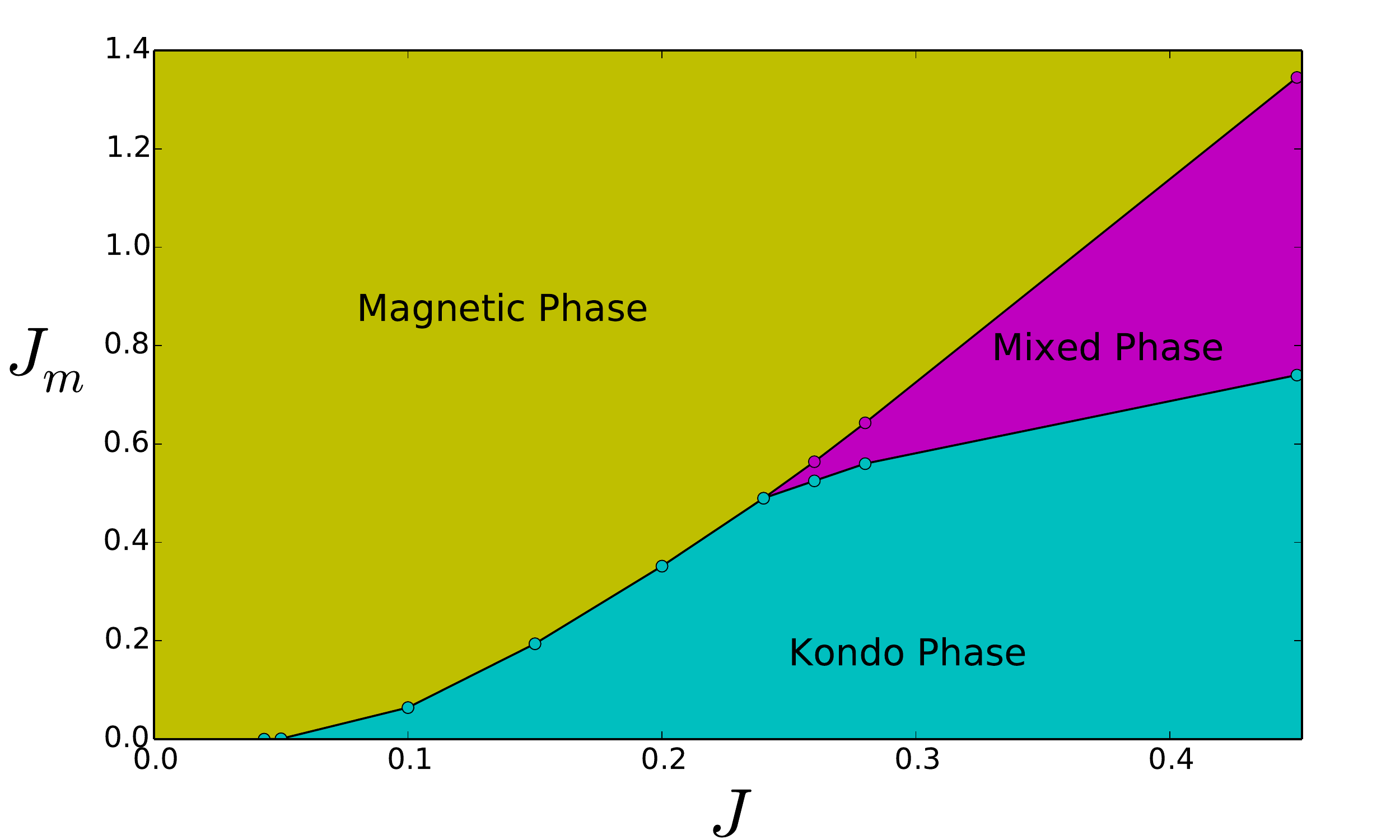} 
{\bf c)}\includegraphics[width=0.30\textwidth]{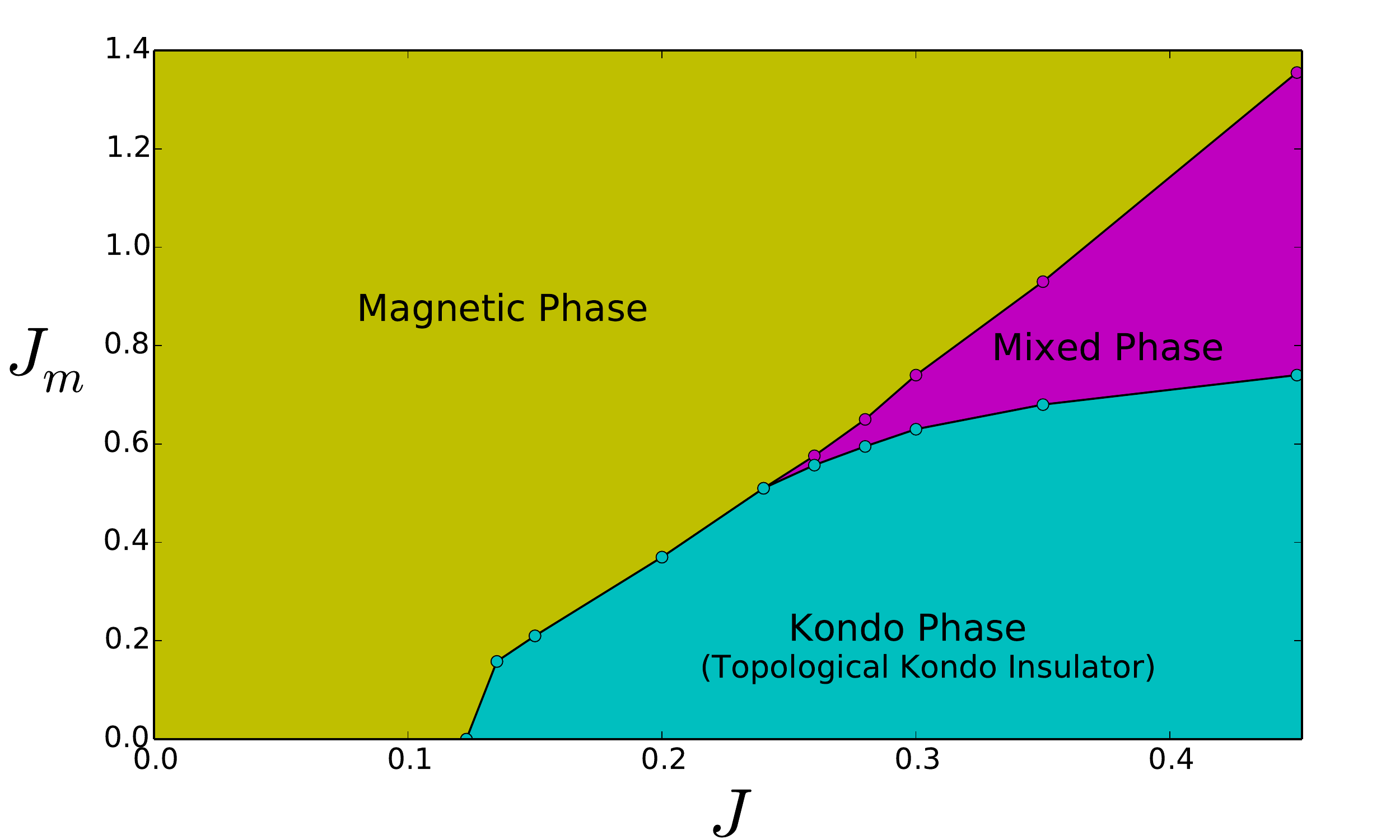}
\caption{ Cuts showing the 3D phase diagram.  {\bf a)} The $J-\Lambda$ plane when $J_m=0$.  {\bf b)} The $J-J_m$ plane when $\Lambda=0$. {\bf c)} The $J-J_m$ plane when $\Lambda=0.5$.  See text for the description of the three phases.
\label{cut1}}\end{figure*}

We verify these results is Fig.~\ref{order1}, through the method detailed earlier.  We see that our results agree with the general argument just presented. In particular, we see that when $\Lambda = 0$, $\phi$ decreases continuously to zero implying that there is a second order transition at a finite $J_c$ between the phase with no Kondo order and the Kondo phase.

We also consider case where $\Lambda \neq 0$, where we show the numerical data for $\Lambda/t = 0.1$ and $0.5$. As expected, we now see a splitting between the $\phi_+$ and $\phi_-$ order parameters.  We also see that the phase transition becomes first order.  In fact we can always find a self-consistent solution for any value of the order parameters down to $\phi_\pm=0$, but as shown in Fig.~\ref{order1}, these solutions are of higher energy than the disordered phase.  This is easy to understand, as a finite $\Lambda$ lowers the variational energy of the fully polarized phase more than it lowers the energy of the Kondo phase.  This is because $d$-electron-like bands in the Kondo phase are necessarily linear combinations of both $d_+$ and $d_-$ electrons, while in the fully polarized phase every d-site is filled with $d_-$ electrons.  This causes an energy crossing between the two possible phases, which occurs away from the critical point. This pushes the fully polarized phase into Kondo regime as you increase $\Lambda$, and causes a first order transition.  Fig.~\ref{cut1} a) shows the phase diagram in the $J-\Lambda$ plane for $J_m=0$.  The transition at $\Lambda=0$ is continuous with the critical point described above, while for $\Lambda \neq 0$, the transition is driven first order.

\subsection{Magnetic Order}
\label{sec:magnetic-order}

Next, we ask what happens when we include the spin-spin interactions between the $d$-electrons.  In this case, there is an interplay between three competing forces, the spin-orbit coupling, the Kondo interaction and the magnetic interaction.  Despite this competition, there exists a phase in which both Kondo and magnetic orders coexist. This is characterized by a nonzero value of both $\langle C^\dagger d \rangle$ and $\langle S \rangle$.  Here we will look at the stability of the Kondo phase to both the fully polarized phase and this mixed order phase.

First consider the fully interacting Hamiltonian.  Deep within the Kondo phase, the mean field solution to $H^{MF}(\langle C^\dagger d \rangle \neq 0 )$ is a saddle point of the action. When $\Lambda \neq 0$, there is a band gap separating the highest occupied band from the lowest unoccupied band so that there are no gapless excitations.  We can then treat the spin-spin interaction as a perturbation around this solution, but it will obviously have no effect if the interaction strength is smaller than the gap.  At the mean field level, the energies of some occupied states will shift down while an equal number of states will have their energy shifted up. Since all states are a finite energy below the Fermi level due to the band gap, all these states will remain occupied and to lowest order in perturbation theory there will be no effect.  

The situation is slightly less obvious when $\Lambda =0$ and the band structure is gapless. In the Kondo phase the bands touch at the Fermi level only at the $\Gamma$ point in the Brillouin zone. This gapless band touching is guaranteed by the fact that $V_\pm(\vec{k}=0)=0$.  In this case, a perturbation on the $J_m=0$ MF solution will raise the energy of some states near $\vec{k}=0$ above $\epsilon_F$.   The change in energy due to this shift is given by $\sim J\langle S \rangle$, while the number of states is limited to $\sim J \langle S \rangle N(\epsilon_F)$.  Meanwhile, from $\mathcal{H}^{MF}$, such a perturbation has a constant energy cost of $J_m\langle S \rangle^2$.  To first order, the total change in energy would be
\beq
\Delta E \sim - c_1 (J_m\langle S\rangle )^2 N(\epsilon_F) +  J_m (\langle S \rangle)^2.
\eeq
Since $N(\epsilon_F)\rightarrow 0$ , it seems likely that even in the presence of gapless states near the $\Gamma$ point, a perturbation in $J_m$ will have a very small effect on the Kondo phase. We verify both these claims by explicit calculation.

Fig.~\ref{order1} b) shows this behavior for a cut in the phase diagram along $J=\text{const}$ and $\Lambda = \text{const}$.  For $J_m$ small, there is no magnetic order and the Kondo phase is stable.  If the gap to single spin excitations is smaller than the energy cost of destroying the Kondo phase, then the mixed phase is stable for some regime of $J_m$.  The numerical calculation shows the spin order parameter $\langle S \rangle$ turning on for some finite $J_m$ and coexisting with the Kondo order parameter $\phi$. As the spin-order increases there is a corresponding drop in the Kondo order parameter.  At some points in phase space, for small $J$, the energy of the fully polarized phase is always lower than the energy of the mixed phase.  In this case, there is a direct transition from the Kondo phase to the magnetic phase.  The energy crossings in Fig.~\ref{order1} b) show that the transitions both into the mixed phase and into the fully polarized phase are first order.

\subsection{3D Phase Diagram}
\label{sec:3d-phase-diagram}

Fig.~\ref{cut1} shows 2D cuts of the phase diagram at $J_m=0$ and at two different values of $\Lambda$. Taken together these gives us the full 3D mean field phase diagram of our interacting model.

When $J_m=0$, there is no competing magnetic interaction.  At $\Lambda=0$ there is a second order transition into the Kondo phase at $J=J_c$. As discussed, this critical point is a result of the vanishing density of states near $\epsilon_F$. In the non-Kondo phase  the $d$-electrons have no order but are unstable to any infinitesimal interaction. For $\Lambda\neq0$, the $d$-levels are split and a full gap opens in the Kondo band structure which gives the phase a nontrivial topology. Here, the transition between the polarized phase and the Kondo phase is driven first order.

We also show cuts in the $J-J_m$ plane along $\Lambda=0$ and $\Lambda=0.5$.  The main difference between these is that when $\Lambda \neq 0$, Kondo phase is destroyed suddenly at small $J$.  Further, when $\Lambda\neq 0$ the Kondo phase has a full band gap with a nontrivial topology, while the Kondo phase for $\Lambda=0$ is a semimetal.  The similarity of the two cuts is due to the stability of the Kondo phase against $J_m$ in both cases.  At the points in phase space where there is nonzero magnetization, $J_m$ is generally large enough that the effect of $\Lambda$ only slightly moves the boundaries.

\section{Edge states of the TKI phase}
\label{sec:edge-states-tki}

Perhaps the most dramatic consequence of symmetry protected topological states is the necessary existence of nontrivial edge states in systems with a boundary.  In two dimensions the allowed edge states are either A) gapless or B) spontaneously break the symmetry, while the system remains gapped with unbroken symmetry within the bulk \cite{PhysRevB.86.125119}.  In three dimensions a third allowed possibility is a surface with topological order \cite{PhysRevB.88.115137,2013arXiv1306.3286M,PhysRevB.89.165132}.  It would be very interesting if such a surface state could be achieved with a topological Kondo insulator, however in our 2D system we must restrict ourselves to the first two possibilities.  It has been shown that on the surface of 3D TKIs, fluctuations around the mean-field state can lead to strongly interacting surface theories  \cite{PhysRevB.90.081113}.  We take a similar approach below for 2D TKIs where we first discuss the mean-field solution and then go beyond MFT to look at the true low energy theory of our edge states.

\subsection {\bf Mean Field Analysis}
\label{sec:mean-field-an}

\begin{figure}[b]
\includegraphics[width=0.8\columnwidth]{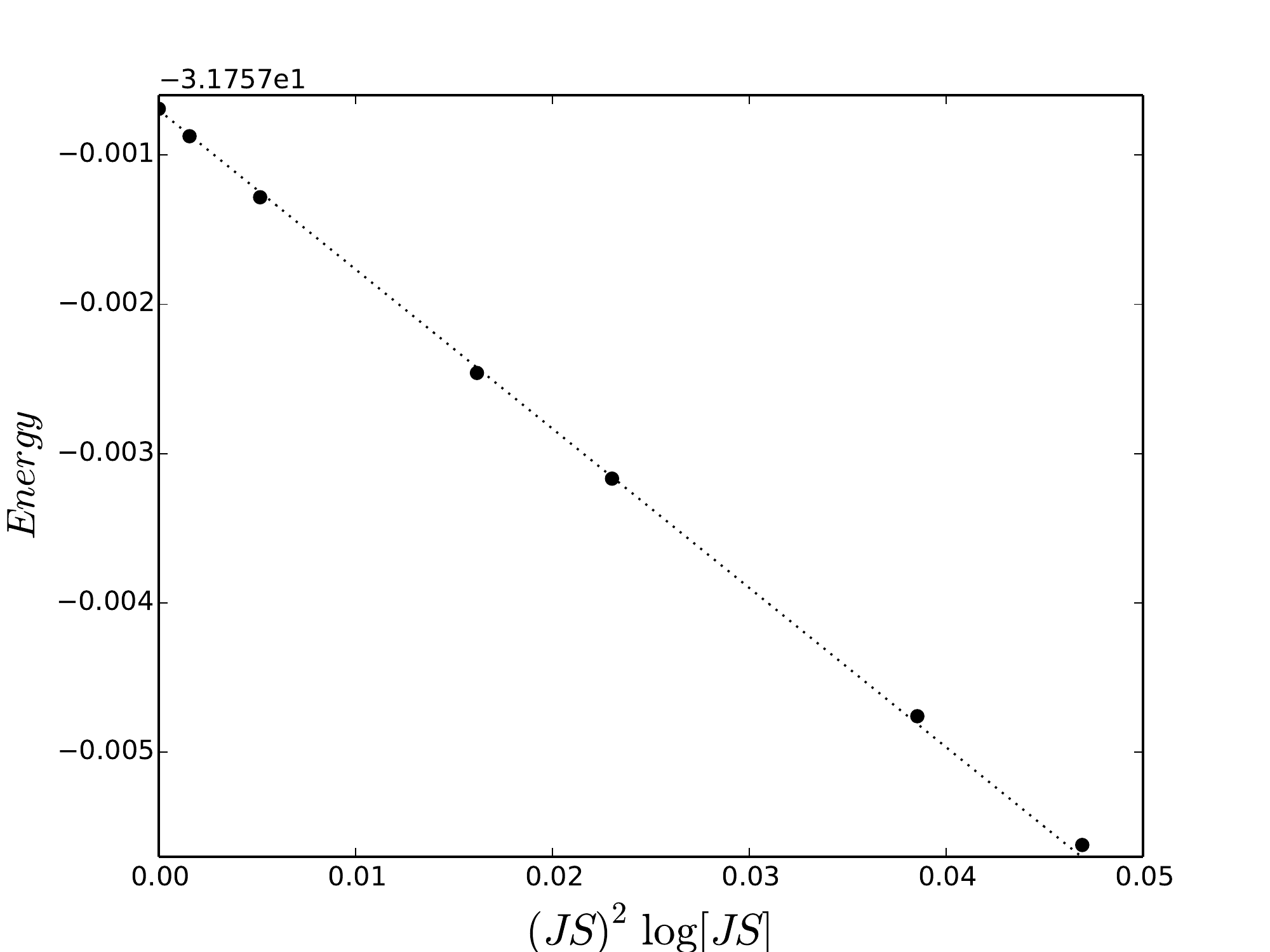}
\caption{The energy gained by opening a gap at the Fermi level near $k=\pi$ is logarithmic in $JS$ in the mean-field calculation. \label{edgeMF1}}
\end{figure}

\begin{figure}[t]
\includegraphics[width=0.49\columnwidth]{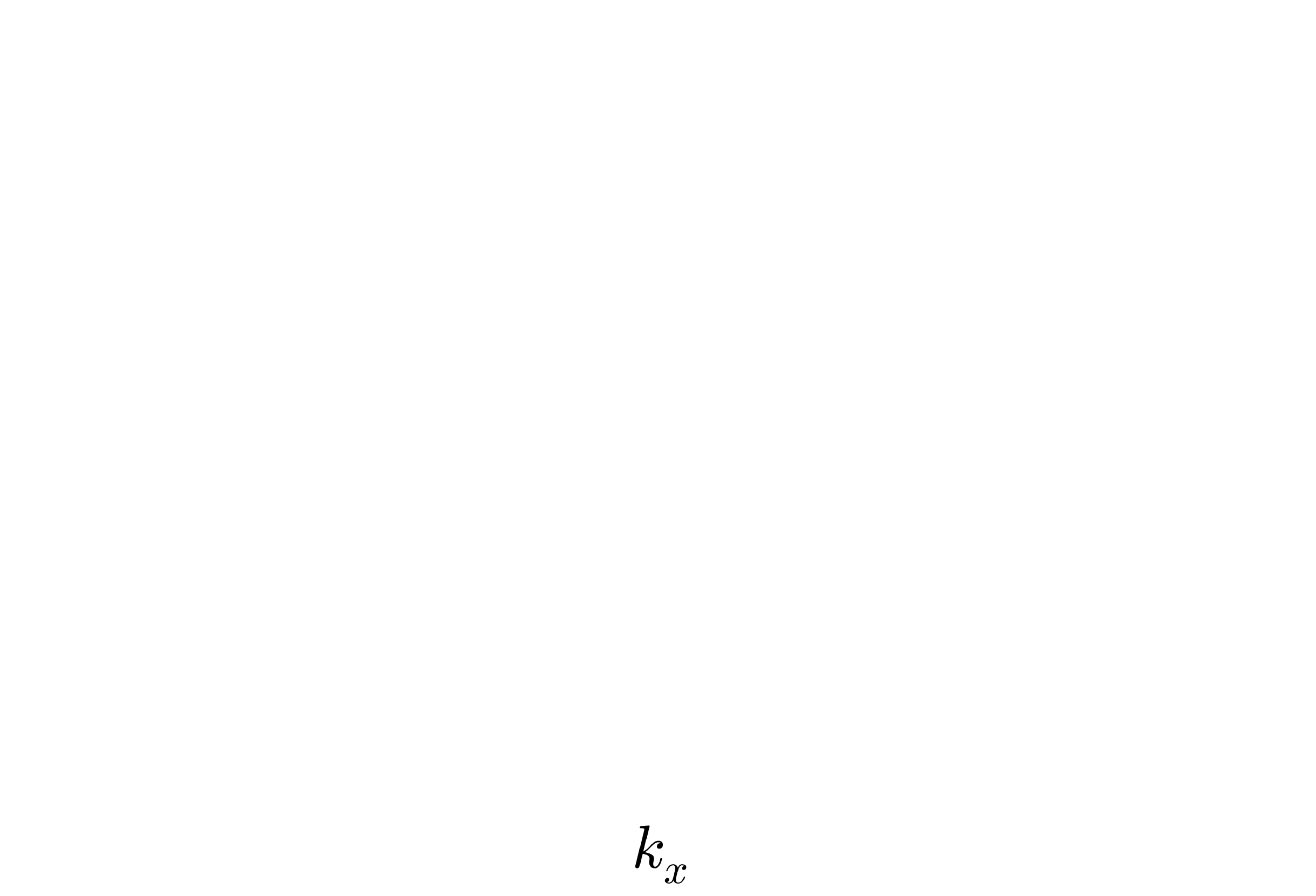}
\includegraphics[width=0.49 \columnwidth]{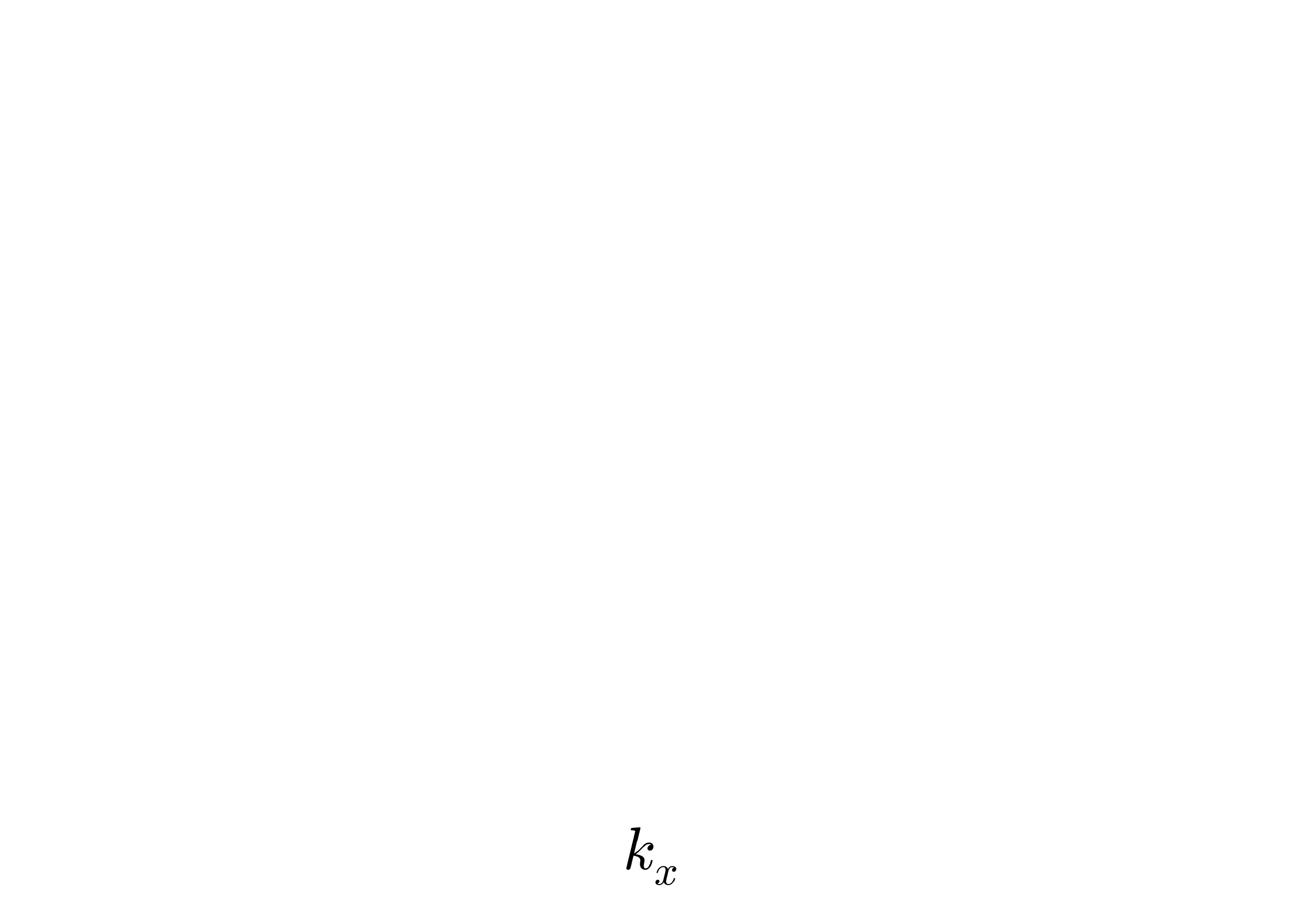}
\includegraphics[width=1.0\columnwidth]{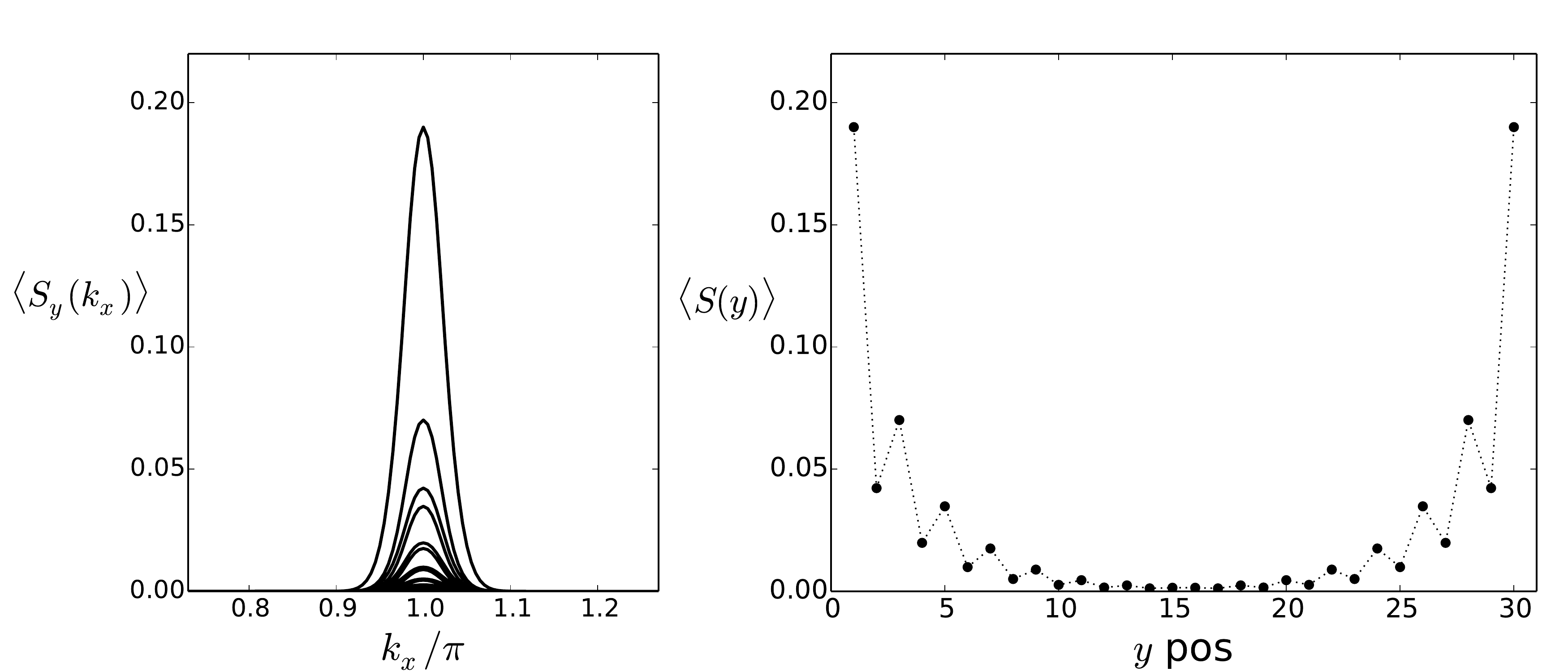} \\
\caption{ {\it(top)} The single-particle band structure on a finite strip shows in gap edge states and a gap opening in the edge spectrum for small $J_m$. {\it (bottom)} Measuring the symmetry breaking shows that the magnetic order is localized near $k_x=\pi$  {(\it left) }, and at the edges of the strip $y=0,L_y$ {\it (right)}. ($y=0$ and $y=L_y/2$ gives the largest and smallest peaks respectively.)
\label{edgeMF2}}\end{figure}

We now show that at the mean field level, when the system is studied on a finite strip,
 so that the noninteracting TKI phase contains gapless edge states,
 the edges are always unstable to magnetic perturbations at the Fermi level. 
In this way, the TKI is a simple way to realize a time-reversal invariant insulator with spontaneous breaking of the TR symmetry on the edges.  

In the bulk system, deep within the Kondo regime, 
the effect of an infinitesimal RKKY interaction is negligible due to the presence of a gap to any spin-1 excitations.

Our argument for the MF edge theory is similar to the Peierls argument \cite{peierls1955quantum, Rice1973125, giamarchi2003quantum} whereby in one dimension, if we ignore the dynamics of the phonons, 
logarithmically more energy can always be gained by the electrons ordering. 
The mean  field $\langle S \rangle$ plays the role of the static phonon fields,
 and the periodic lattice distortion is instead replaced by a coupling between right and left moving electron fields.
 The effect of the magnetic interaction is to open a gap at the Fermi level, replacing the edge spectrum
 $\epsilon_k \sim v_k k$ with  $\epsilon_k \sim \sqrt{(v_k k)^2 + \Delta^2}$ where $\Delta = J_m \langle S \rangle$.  
Following the standard argument, one dimensional gapped systems contain a singularity in the density of states, and a proper 
estimate for the change in energy due to the magnetic ordering is 
\beq
\delta E \sim \int_0^\Lambda dk(\sqrt{(v_k k)^2 + \Delta^2} - v_k k )  = \frac{1}{4} \Delta^2 \log[4 e \Lambda^2/\Delta^2].
\eeq

We verify this argument with an explicit mean-field calculation in the same vein as in section II for the bulk phase diagram.  We perform the mean-field calculation in the same way as for the bulk system. The Hamiltonian now has an $8N_y$ site basis, where $N_y$ is the number of unit cells of our finite strip in the $y$ direction. We use different order parameters for sites on the edge and in the bulk and find the self-consistent values of these order parameters. For the bulk order parameter we average the order over all sites that are not on the edge.  In Fig.$\ref{edgeMF2}$, we have tuned $J$ and $\Lambda$ to a point deep within the Kondo phase.  We saw in the previous section that this phase is extremely robust against the formation of magnetic order.  An important point is that in order to satisfy the conditions $\langle d^\dagger d \rangle = \langle c^\dagger c \rangle = 2$ on every site, the edge states need to intersect the Fermi level, $\epsilon_F$, at exactly $k=\pi$. This is the time-reversal invariant momentum at which, due to Kramer's theorem, the two edge states intersect.  In the mean field, the spin-spin interaction term couples fermions of opposite spin at the same momentum. At $k=\pi$, these two fermions modes are degenerate with energy $\epsilon = \epsilon_F$. The spin interaction breaks this degeneracy, sending one state above $\epsilon_F$ and one below $\epsilon_F$.  Consider the Hamiltonian describing of the edge states near $k_F$,
\beq
H_{\text{edge}} = \sum_{-\Lambda<k-\pi<\Lambda} \left( \begin{array}{c} c^\dagger_{k\uparrow} c^\dagger_{k \downarrow} \end{array} \right) \left[\begin{array}{cc}\epsilon_k & J_m \\  J_m & -\epsilon_k  \end{array} \right] \left( \begin{array}{c}c_{k \uparrow} \\c_{k \downarrow} \end{array}\right).
\eeq
The eigenvalues of $H$  are $\epsilon_{\pm} = \sqrt{\epsilon_k^2 + J_m^2}$.  When $\epsilon_k \ll J_m$, the corresponding eigenstates are given by $v_{\pm} = c_{k\uparrow} \pm c_{k \downarrow} + \mathcal{O}(\frac{\epsilon}{J})$.  Meanwhile, for $J\ll \epsilon_k$, to first order the eigenstates are just the original states $v_+ = c_{k\uparrow} + \frac{J}{2\epsilon_k} c_{k \downarrow}$ and $v_- = c_{k \downarrow} - \frac{J}{2 \epsilon_k} c_{k \uparrow}$. In the first case, the conduction electrons are almost completely polarized in the XY plane, while in the second case the eigenstates have very little magnetic order. Therefore, it is only in the regime where $\epsilon_k \ll J_m$, that a small $J_m$ creates a significant magnetization.  But since the edge states are gapless there will always be some finite region where this condition is true, and it is these small number of states which contribute to the spontaneous breaking of TR symmetry on the edge.

In Fig.~\ref{edgeMF1}, we verify that the energy of a self-consistent solution of the MF Hamiltonian as a function of the input parameter $JS$ is 
\beq
\Delta E \sim (J_m S)^2 |\log(J_m S/\tilde{\Delta})| - J_m S^2
\eeq
where $\Delta$ is an arbitrary cutoff.  Therefore, for any value of $J_m$, there is an $S$ for which the energy gain is positive, making this TR broken state favorable.  Since the self-consistent point $S=\langle S \rangle$ is the variational minimum, it must therefore also have a positive gain in energy.  Fig.~\ref{edgeMF2} c) and d) show that the symmetry breaking is localized to momenta near the TRIM $k=\pi$ and is localized in space to the edge of the strip. Meanwhile, Fig.'s \ref{edgeMF2}  a) and b) show the band structure of the finite strip.  In particular, they show the opening of a gap at the Fermi level due to a small nonzero $J_m$.

\subsection{Luttinger Liquid Physics} 
\label{sec:lutt-liqu-phys}

The Peierls argument and mean-field calculations in the previous section are only valid in the limit of a static spin field (similar to how the lattice Peierls argument is only valid for static phonons). For a one-dimensional system, however, we can solve the low-energy theory exactly using bosonization techniques. In gapless 1D system, the low-energy excitations are bosonic degrees of freedom which are localized near the Fermi points. For a non-interacting topological insulator it is well known that each edge contains a time-reversed  pair of chiral fermion modes \cite{PhysRevLett.95.226801}, for example a right-moving spin up mode and a left-moving spin-down mode.  Taken together these comprise a single bosonic degree of freedom, so that the problem maps onto a spinless fermion problem. In the absence of interactions, which may couple the right and left moving modes, the bosonic action for the edge of our TKI is just the action of a free bosonic particle
\beqa
S = v_k \int dx d\tau \left [ K (\partial_x \phi)^2 + \frac{1}{K} (\partial_\tau \phi)^2 \right] \label{eqn:boseaction}.
\eeqa

In our case, the neutrality condition
\beqa
\frac{1}{N} \sum_k \langle d^\dagger d \rangle = \frac{1}{N} \sum_k\langle c^\dagger c \rangle = 2 
\eeqa
ensures that the edge states cross the Fermi level at exactly $k = \pi$.  Near this Fermi level, the right and left moving modes can be bosonized to give a single free bosonic degree of freedom which describes the low-energy dynamics of the system.   This allows us to give a rather simple interpretation of the mean-field result in terms of the allowed interaction between the low-energy modes of our theory. We will further see how the mean field result fails to properly capture the fluctuations of the 1D edge modes.  

At the Fermi point we have a single right moving spin up mode and a left moving spin down mode. The spin-spin interaction, $J_m$, couples these modes together.  We can expand the fermion creation and annihilation operators in terms of bosonic operators.  The most relevant terms in this expansion are
\beqa
\psi^\dagger_{R/L} \sim e^{i (\theta \pm \phi)}.
\eeqa

At the mean field level, the spin-spin interaction is $J_m \langle S \rangle S^x$, where $S^x = (d^\dagger_{+ \uparrow} d_{+\downarrow} + d^\dagger_{- \uparrow} d_{- \downarrow} + \text{h.c.})$. $\psi_R$ is a linear combination of all spin up $c$ and $d$ operators, and likewise for $\psi_L$ and spin down operators.  Therefore, writing this interaction in terms of $\psi_{R/L}$ produces a number of terms. The most relevant allowed terms is,
\beqa
g_{_{\text{MF}}} \left (\psi_R^\dagger \psi_L + \psi_L^\dagger  \psi_R \right )\sim \cos(2 \theta),
\eeqa
where $g_{_{\text{MF}}} \sim J_m \langle S \rangle$. The scaling dimension of such an operator is well known from the form of the free bosonic correlation function to be
\beqa
\dim[\cos(p\theta)] = \frac{p^2}{4 K} \hspace{1mm}, \hspace{1mm} \dim[\cos(p\phi)] = \frac{p^2 K}{4} . \label{a} 
\eeqa
Therefore the operator $\cos(2 \theta)$ is a highly relevant perturbation at the noninteracting point $K=1$.  This implies that for the mean-field model, an infinitesimal perturbation $J_m$ will flow under RG to strong coupling.  The resulting model is a sine-Gordon model, in which the $\theta$ field is pinned at strong coupling, breaking TR symmetry. 

However, this type of naive analysis ignores the fluctuations of the spin order parameter $\langle S \rangle$, which can drastically affect the physics. In particular, we can write down all allowed interactions involving a single right and a single left moving mode.  These interactions are
\beqa
H_{\text{int}}^{(1)} &=& g_1\psi_R^\dagger \psi_R \psi_L^\dagger \psi_L  \label{bose1} \\
H_{\text{int}}^{(2)} &=& g_2\psi_R^\dagger \psi_R^\dagger \psi_L \psi_L \label{bose2}
\eeqa
where the umklapp operator $g_2$, can only exist in systems with a single fermion species if there is point splitting $g_2 \sim  \psi_R^\dagger(x) \psi_R^\dagger(x+a) \psi_L^\dagger(x) \psi_L(x+a)$. This introduces a derivative upon taking the continuum limit, making this operator less relevant. 

The most relevant bosonized expressions contained in Eq.'s~(\ref{bose1}) and (\ref{bose2}) are
\beqa
H_{\text{int}}^{(1)} &\sim& g_1 \left [ (\partial_x \phi)^2 - (\partial_x \theta)^2 \right ] \\
H_{\text{int}}^{(2)} &\sim& g_2 \cos(4 \theta).
\eeqa
The $g_1$ term can be absorbed into the action, Eq.~(\ref{eqn:boseaction}), by renormalizing the value of the Luttinger parameter $K$. The $g_2$ term, however, will attempt to pin the field $\theta$, thus opening a gap in the energy spectrum at the Fermi level.  This process is relevant if the scaling dimension for the $g_2$ operator is less than $2$.  In that case, the cosine operator will flow to strong coupling and pin the $\theta$ field. By Eq.~(\ref{a}), the scaling dimension of the $g_2$ operator is $\dim[g_2] = 4/K$.  This implies that at the noninteracting point $K=1$, the cosine term is irrelevant and the action of the edge states remains gapless.  It is only when the $g_1$ term pushes the value of $K$ past $K=2$ that the cosine operator becomes relevant.  At this point there is a Kosterlitz-Thouless transition into a phase where the $\theta$ field is pinned.  The $\theta$ field in this case acts like the $S^x$ spin operator since, by the bosonization rules above
\beqa
S^x \sim \psi_\uparrow^\dagger \psi_\downarrow + \psi_\downarrow^\dagger \psi_\uparrow = \psi_R^\dagger \psi_L + \psi_L^\dagger \psi_R = \cos(2 \theta).
\eeqa
Pinning $\theta$ at $\theta=0$, implies that $\langle S^x \rangle \neq 0$, and so time reversal symmetry is broken on the edge. 

The stability of edge states in the spin-quantum Hall effect has been studied previously, focusing on the effect of a screened Coloumb interaction \cite{PhysRevB.73.045322, PhysRevLett.96.106401}.  The results in those works similarly find that a single Kramer's pair of edge modes are stable at weak coupling, but may be driven into a gapped phase by sufficiently strong interactions. In our problem, the analogous interactions are included naturally in the form of the RKKY interaction term.  

This result leaves two possibilities for the full phase diagram in the presence of edge states.  The first case is that the strength of RKKY interactions, $J_m$, required to gap the edge modes for a given value of $J$ is less than that required to drive the edge into a magnetic phase.  In this case, time-reversal symmetry will be broken spontaneously on the edge of the system while being preserved within the bulk, and the edge properties of the TKI in this phase will differ dramatically from that of the uncorrelated 2D topological insulator of Ref's~[\onlinecite{PhysRevX.1.021001},\onlinecite{PhysRevLett.95.226801}].  

The second possibility is that for all $J$, the required $J_m$ to drive the edge to a gapped state is greater or equal to that required to drive the bulk into the magnetic or mixed phases. In this case, the low energy theory of the TKI is qualitatively similar at weak coupling to the noninteracting TKI phase of Ref.~[\onlinecite{PhysRevX.1.021001}].  However, even in this case, the low-energy edge theory, while gappless, is still governed by the action in Eq.~(\ref{eqn:boseaction}) and controlled by the Luttinger parameters $v_F$ and $K$ and thus will show quantitative differences from the noninteracting theory.  

We may also ask, in what case is the mean-field analysis of the previous section valid.  If we generalize our model to one with $N_f$ flavors of fermions on each lattice site we enter a regime where the large-$N_f$ and slave-boson approaches to the Kondo problem become justified \cite{PhysRevB.30.3841,PhysRevB.85.125128}. In the limit $N_f\rightarrow \infty$, the mean-field result becomes exact as fluctuations, even on the edge, are strongly suppressed. Within the Luttinger liquid framework, we now need to study a system with $N_f$ time-reversed pairs of gappless modes on each edge.  In the topologically nontrivial phase, $N_f$ must be odd. It is mentioned in Ref.~\onlinecite{PhysRevB.73.045322} that the $N_f=3$ case is less stable to interactions than the $N_f=1$ case. This is due to the enlarged number of allowed interactions which appear when we include terms which couple different modes together. We therefore spectulate that as $N_f \rightarrow \infty$, the huge number of allowed interactions in the low-energy picture will always gap out all edge modes in order to agree with the high-energy mean field picture.  The large-$N_f$ theory should always fall under the first case discussed above where TR symmetry is broken on the edge but preserved within the bulk.

\section{Conclusions}
\label{sec:conclusions}

In this paper, we have discussed the role of interactions in the physics of 2D topological Kondo insulators.  To this end, we wrote down a realistic theory for a system of graphene doped with 5d adatoms where there are two localized $d$-electrons on each adatom which form a composite spin-1 magnetic moment and interact with the nearby conduction electrons on the graphene lattice.  We performed a mean-field calculation on this model where we were careful to note that included within the Kondo interaction is an RKKY type spin-spin term between the localized $d$-electron states, which appears at second order in perturbation theory in this interaction.  Including both the Kondo and RRKY couplings as separate parameters allows us to decouple our theory into both the Kondo and magnetic ordering channels. 

First, we found that the bulk phase diagram is separated into three distinct regions. For sufficiently large values of the Kondo coupling and weak RKKY interactions the Kondo insulator phase is stabilized and posseses the same low-energy theory as a noninteracting topological insulator.  As the relative strength of the RKKY term is increased the localized moments break time-reversal symmetry by either becoming fully polarized or entering a mixed phase with both magnetic and Kondo order. The nontrivial topological distinction of the TKI phase is completely lost in both of these magnetic phases.

The TKI phase is fully gapped and so we expect that any fluctuations around the mean-field result can be neglected.  Therefore, to look for nontrivial correlation phenomena we further studied the effects of the Kondo interaction on the gappless edge states of this topological phase.

We find that in the large $N_f$ limit, where the mean field treatment becomes exact, a gap is always opened in the edge spectrum and TR symmetry is spontaneously broken locally at the edges.  At weak coupling, when the RKKY parameter $J_m$ is small, there exists a critical $1/N_f$  which marks a qualitative change in behaviour whereby strong fluctuations at the edge destroy the magnetic order of the MF solution.  For $N_f=1$, as in our original microscopic model, this leads to a phase with gappless edge states which are described by the Luttinger liquid Hamiltonian with interaction dependant Luttinger parameter $K$. This 2D TKI therefore provides a natural system where interactions cannot be ignored if one is to properly describe the edge phenomena quantitatively.  At intermediate values of the coupling $J_m$, an edge transition into the gapped phase occurs independantly of the magnetic transition in the bulk which may lead to a qualitative deviation from the noninteracting edge theory.  Indeed, there two possibilities within the TKI phase at intermediate coupling. The first is that the edge states remain gappless everywhere in this phase.  The second is that there is a transition into a phase with magnetic order on the edge but not in the bulk. While we cannot rule out either of these two cases within our analysis, the second possibility represents a dramatic departure from the edge theory of the noninteracting topolgical insulators and thus is a possible avenue to search for correlation effects in topological phases.

{\em Acknowledgements --} This work was supported by the U.S. Department of Energy, Office of Science,
Basic Energy Sciences, under Award No. DE-FG02-08ER46524 and by NSERC of Canada (J.I.).  It benefitted from the facilities of the KITP, supported by National Science Foundation under grant No. PHY11-25915.

\bibliography{base}

\end{document}